\documentclass[useAMS,usenatbib]{mn2e}
\usepackage{graphicx}

\title[Low-$Z$ AGB stars]{The evolution of low-metallicity asymptotic giant
branch stars and the formation of carbon-enhanced metal-poor stars}
\author[H.~B. Lau, R.~J. Stancliffe \& C.~A. Tout]{Herbert H.~B.
Lau$^{1,2,3}$\thanks{hlau@kiaa.pku.edu.cn}, Richard J. Stancliffe$^{2,3}$ 
and Christopher A. Tout$^2$\\
$^1$Kavli Institute for Astronomy \& Astrophysics, Peking University, China\\
$^2$Institute of Astronomy, The Observatories, Madingley Road, Cambridge CB3 0HA\\
$^3$Centre for Stellar and Planetary Astrophysics, School of Mathematics,
Building 28, Monash University, Clayton VIC 3800, Australia}
\begin{document}
\bibliographystyle{mn2e}

\date{Accepted 2009 March 12.  Received 2009 March 12; in original form 2009 February 16}

\pagerange{\pageref{firstpage}--\pageref{lastpage}} \pubyear{2009}

\maketitle

\label{firstpage}

\begin{abstract}
We investigate the behaviour of asymptotic giant branch (AGB) stars
between metallicities $Z=10^{-4}$ and $Z=10^{-8}$.  We determine
which stars undergo an episode of flash-driven mixing, where
protons are ingested into the intershell convection zone, as they
enter the thermally pulsing AGB phase and which
undergo third dredge-up.  We find that flash-driven mixing
does not occur above a metallicity of $Z=10^{-5}$ for any mass of
star and that stars above 2$\,\rm M_\odot$ do not experience this phenomenon at
any metallicity.  We find carbon ingestion (CI), the mixing of carbon into
the tail of hydrogen burning region, occurs in the mass range 2$\,\rm M_\odot$
to around 4$\,\rm M_\odot$.  We suggest that CI may be a weak version
of the flash-driven mechanism.  We also investigate the effects of
convective overshooting on the behaviour of these objects.  Our models
struggle to explain the frequency of CEMP stars that have both significant
carbon and nitrogen enhancement.  Carbon can be enhanced through
flash-driven mixing, CI or just third dredge
up.  Nitrogen can be enhanced through hot bottom burning and the
occurrence of hot dredge-up also converts carbon into nitrogen.  The C/N
ratio may be a good indicator of the mass of the primary AGB stars.
\end{abstract}


\section{Introduction}

In recent years, there has been considerable interest in the
observation of metal-poor stars and particularly those of a
carbon-rich nature.  A surprising fraction of stars with metallicities
below [Fe/H] of around $-2.5$ are found to be rich in carbon.  This
fraction, while uncertain, may be as high as around 20\,per cent
\citep{2006ApJ...652L..37L}.  Many of these stars display evidence of
$s$-process enrichments.  It has been proposed that they may be formed
by mass transfer in binary systems.  Such a system would consists of
two stars in a wide orbit.  The primary star evolves to the asymptotic
giant branch (AGB) where it becomes enriched in carbon (and also
$s$-process elements).  Strong mass loss strips the envelope from this
star which becomes a white dwarf and fades from view.  The star is
losing mass so some of this is accreted by the companion which becomes
carbon-rich.  It is this companion that we see today.  Credence is
lent to this argument by the observation that, of the carbon and
$s$-process rich stars, around 70\,per cent are found to be members of
binary systems \citep{2005ApJ...625..825L}.  This is consistent with
the whole population being binaries.

This scenario may be complicated somewhat.  It has commonly been
assumed that, when the material is accreted on to the companion, it
remains on the surface of the recipient star.  However,
\citet{2007A&A...464L..57S} showed that thermohaline mixing can lead
to this accreted material being extensively mixed with the pristine
matter of the companion and this can have important consequences for
the abundances that would be observed in such an object.  The extent
of this mixing depends on the composition and mass of accreted
material, as well as the evolutionary state of the accreting star
\citep[see][]{2008MNRAS.389.1828S}.  However, we may still hope to
observe the signatures of low-metallicity AGB stars in the
compositions of the CEMP stars that are still visible today.  To do
this demands that we know what the expected nucleosynthetic signatures
of the parent AGB stars are.

The asymptotic giant branch (AGB) is a late phase of stellar evolution
for stars with masses from 1 to around 8$\,\rm M_\odot$.  It follows the end of the core helium burning
phase.  As the core of the star runs out of helium it begins to ascend
the AGB.  The star first expands and cools and the convective envelope
deepens.  The envelope can reach down as far as the hydrogen burning
shell, pulling material to the surface and altering the surface
composition of the star.  This is the second dredge-up

Helium and hydrogen are both burning in thin shells.  The helium
burning shell slowly moves outwards in mass and the region between the
two shells narrows.  When they get very close (separated by a few
hundredths of a solar mass) the star enters the thermally pulsing
asymptotic giant branch (TP-AGB) phase.  Thermal pulses are the rapid
increase in the helium-burning luminosity to over $10^{8}\,L_\odot$,
for a brief period of time, usually around 10yr.  The thermal pulses
are a consequence of the thinness of the helium burning shell and high
temperature sensitivity of the triple-$\alpha$ reaction.  The theory
is described, for AGB stars, by
\citet{1965ApJ...142..855S}.  When energy is dumped in the thin shell,
the temperature can rise when the shell expands.  There is no
thermostatic control until the shell is no longer thin.  Also, for the
pulse to occur, the increased radiative losses owing to the increase
of shell temperature must not carry energy away faster than it is
being generated.  This is usually fulfilled because helium burning is
highly sensitive to temperature.
\citet*{2004A&A...425..207Y} have investigated the stability of shell
sources and find that the helium burning shell tends to be more stable
at high temperature because the triple-$\alpha$ reaction is less
temperature sensitive.

During the thermal pulses, the sudden increase in energy release,
caused by enhanced helium burning, leads to convection between the two
burning shells and mixes the ashes of helium burning outwards from the
helium burning shell.  The star expands and the hydrogen burning shell
cools and may be extinguished.  Eventually thermostatic control is
restored in the He-shell, the helium luminosity starts to drop and
the intershell convection region shuts down.  The convective envelope
deepens during this power down phase, which lasts for about $100\,$yr.
The envelope could penetrate into regions where the intershell
convection zone had been active \citep{1975ApJ...196..525I}.  This
means that the carbon that has been produced during helium burning is
ingested by the envelope.  This is called third dredge-up (TDUP).  The
surface is then enriched in carbon (and $s$-process
elements\footnote{The $s$-process is the capture of neutrons by nuclei
in conditions of sufficiently low neutron density that any unstable
products have time to decay before another neutron is captured.  The
neutron source in AGB stars is not yet properly understood.}).  The
base of the convective envelope can be hot enough for some CNO cycle
reactions to occur, particularly for stars more massive than
$4\,M_\odot$.  This is called hot bottom burning (HBB).  Helium
burning returns to a quiescent state and the hydrogen shell reignites.
Helium accumulates again during this interpulse period which can last
over $1{,}000\,$yr.  The cycle then repeats and may do so many times
before stellar winds can strip away the envelope.

Low-metallicity AGB stars are known to evolve differently to their
higher metallicity counterparts.  In the case of the low-mass objects
it is possible for the helium-burning driven intershell convection
zone to penetrate into the hydrogen-rich envelope
\citep{2000ApJ...529L..25F}.  This does not happen at higher
metallicity because the hydrogen burning shell presents an effective
entropy barrier which inhibits the advance of the intershell
convection zone.  As the metal content of the star is reduced,
hydrogen burning via the CNO cycle is weaker and presents less of a
barrier.  Below some critical metallicity, the intershell zone manages
to penetrate into H-rich regions.  This leads to hydrogen being
dragged down to regions of high temperature, with vigorous H-burning
occurring as a result.  The convective region can then split in to two
separate zones, one driven by H-burning, the other by He-burning.
This phenomenon is referred to as flash driven mixing \citep[FDM,
][]{2004ApJ...611..476S}.  It has been reported by several authors,
including \citet{2000ApJ...529L..25F} and
\citet{2008A&A...490..769C}.\footnote{The community is yet to accept a
standard nomenclature for these events.  They have also been called
`proton ingestion episodes' \citep{2007ApJ...667..489C} and `dual
shell flashes' \citep{2008A&A...490..769C}.}

Metal-free evolution for low- and intermediate-mass stars has been
explored by \citet{2001ApJ...554.1159C} and
\citet{2002ApJ...570..329S}.  \citet{2001ApJ...554.1159C} find that
during thermal pulses of AGB stars at or below 5$\,\rm M_\odot$, a
convective shell forms at the H-He interface and the subsequent
expansion of this convective shell in to the underlying layers
previously occupied by the pulse dredges up some carbon and initiates
an H flash.  In the paper, this phenomenon is referred as carbon
ingestion (CI).  \citet{2002ApJ...570..329S} also find this unusual
convective shell and that there is a mixing episode for stars at or below
6$\,\rm M_\odot$.  They also refer to it as CI.

In addition, \citet{2004ApJ...605..425H} reported that, at a
metallicity of $Z=10^{-4}$, the nature of third dredge-up changes.
With the inclusion of convective overshooting, he found that hydrogen
would actually burn {\it during} TDUP and that the propagation of a
H-burning flame could erode the entire intershell and even dredge up
some material from the CO core.  This behaviour has been dubbed hot
dredge-up.

Our purpose in this work is to delineate the boundaries between
these regimes of behaviour.  This will enable us to place limits on
which stars are able to produce which nucleosynthetic signatures.
Based on our models, we can hope to explain the variation of carbon
and nitrogen enrichment of CEMP stars in terms of the different
enrichment processes of extremely metal-poor AGB stars of different
masses.  Therefore, the observed frequencies of different types of CEMP
stars may tell us the relative numbers of low- and high-mass AGB stars
in extremely metal-poor environments.  This can help deduce the form
of the initial mass function at low metallicity \citep[see
e.g.][]{2005ApJ...625..833L}.

\section{Stellar evolution code}

\begin{figure*}
\includegraphics[width=14cm, angle=270]{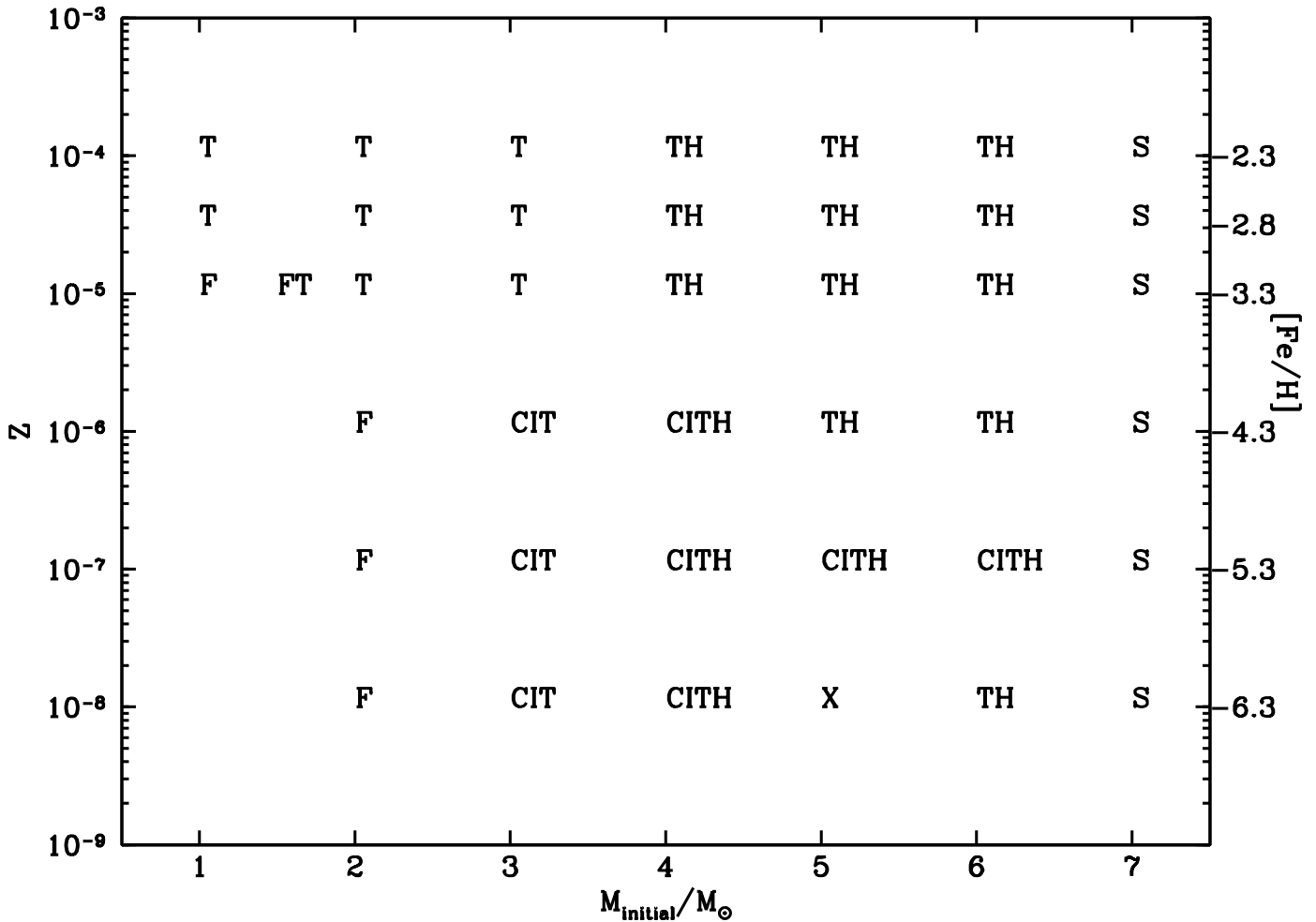}
\caption{AGB behaviour of the models as a function of mass and metallicity without overshooting.
The symbols are T -- the model undergoes third dredge-up, H -- the
model undergoes hot third dredge-up, F -- the model undergoes
flash-driven mixing, CI -- the model undergoes a carbon
injection, X -- no TDUP occurs, S -- the model undergoes carbon
ignition before the AGB and presumably becomes an S-AGB star.}
\label{fig:ZversusM}
\end{figure*}

\begin{figure*} 
\includegraphics[width=14cm, angle=0]{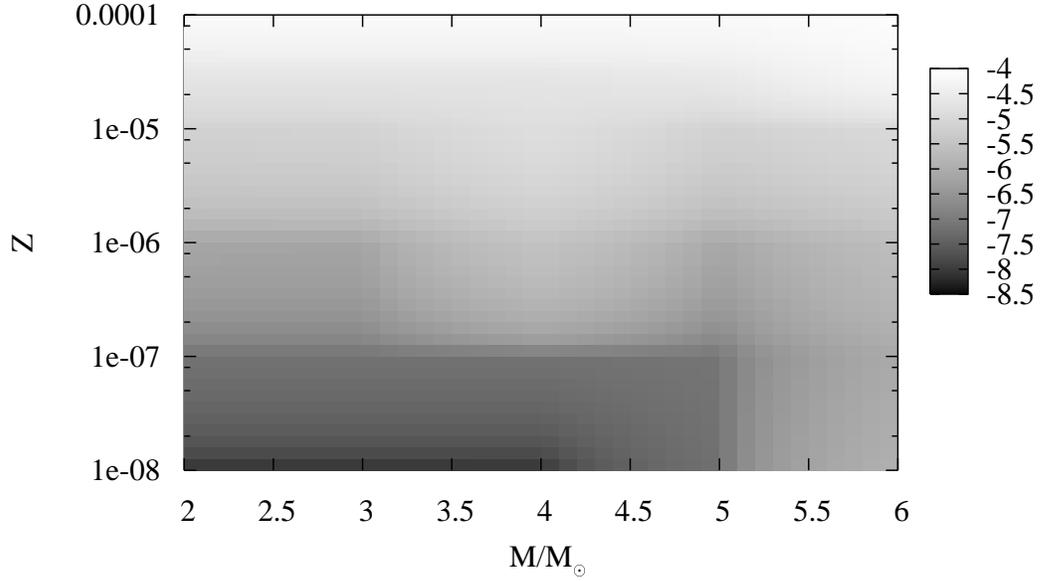}
\caption{The abundance of CNO elements by mass at the beginning of the TP-AGB.
The action of second dredge-up substantially raises the abundance in
the case of the intermediate-mass stars.}
\label{fig:Z2dup}
\end{figure*}

\begin{figure*} 
\includegraphics[width=14cm, angle=0]{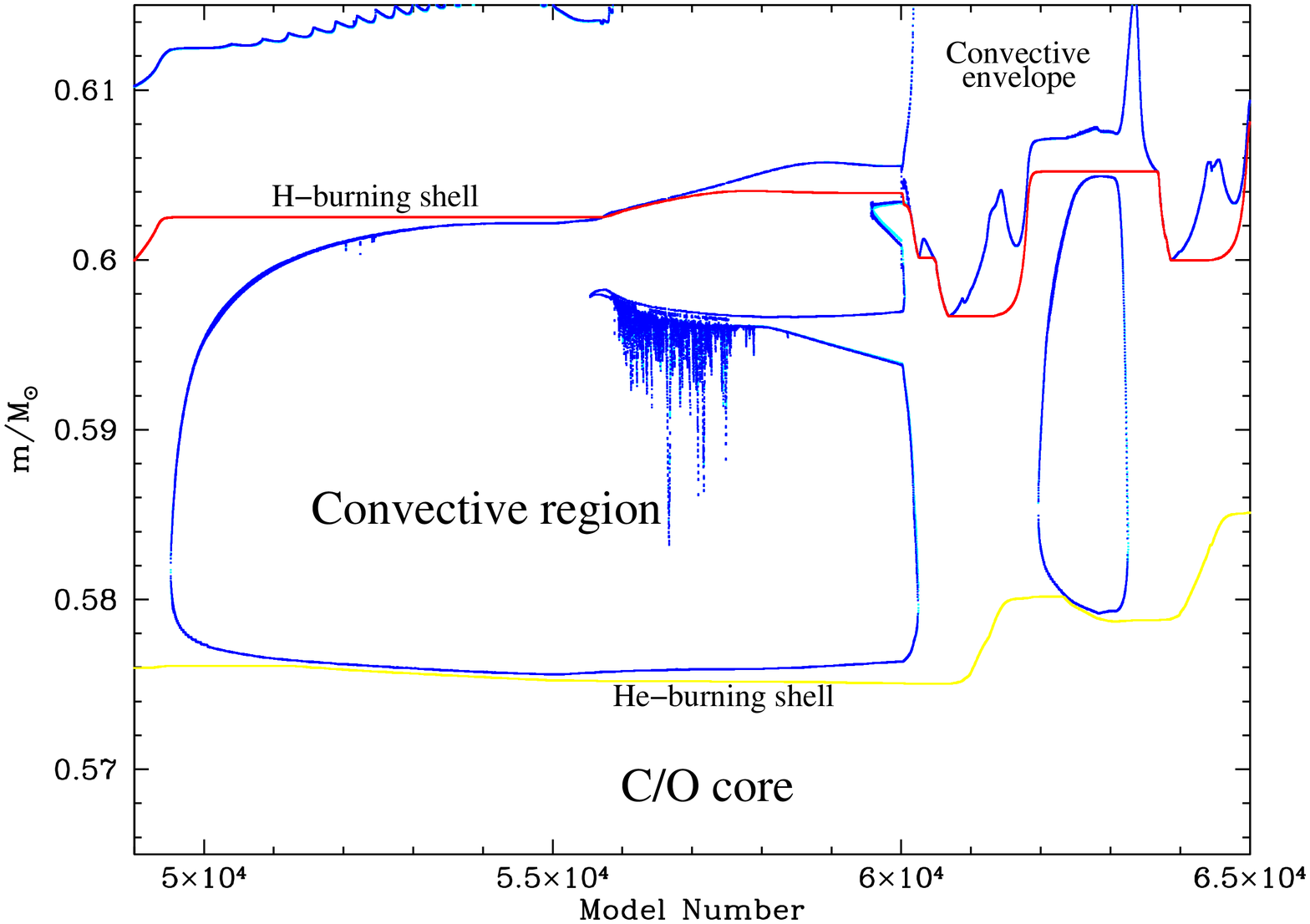}
\caption[Evolution of $1.5\,M_\odot$, $Z=10^{-5}$ model during He-FDM]
{Evolution of $1.5\,M_\odot$, $Z=10^{-5}$ model during He-FDM.  We
can see the convective region reaches the H-burning shell at model
number around 56000.  The messy appearance is due to numerical
problems to resolve the mixing episode.  The 2nd pulse, at about model
$62{,}000$ is simply a normal thermal pulse.}
\label{fig:1.5FDDM}
\end{figure*}

We use the stellar evolution code {\sc stars} which is a variant of that
originally developed by \citet{1971MNRAS.151..351E} and updated by
many authors \citep[e.g.][]{1995MNRAS.274..964P}.  The version used
here includes the AGB-specific modifications of
\citet*{2004MNRAS.352..984S}, together with the updated opacity tables
of \citet{2004MNRAS.348..201E} which account for the enhancement to
the opacity from enhancements in carbon and oxygen abundance.  The
code has been extensively used to compute TP-AGB evolution in recent
years \citep[see e.g.][]{2005MNRAS.356L...1S, 2007MNRAS.378..563L}

We do not apply mass loss at any stage of evolution.  Mass-loss rates
for AGB stars are very uncertain and it is expected that mass loss
scales with metallicity.  In any case, the behaviour we are interested
in happens in the first few pulses before any mass loss would be
significant.  We use $\alpha = 2.0$ for the mixing length parameter in
the \citet{bohmvitnse1958} prescription.  This value is chosen based
on calibration to a solar model.  We also chose solar-scaled initial
abundances \citep{1989GeCoA..53..197A} rather than attempting to
follow the chemical evolution of the Galaxy.

We have made models of metallicity $Z=10^{-4}, 3\times10^{-5},
10^{-5}, 10^{-6}, 10^{-7}, 10^{-8}$ and $0$.  We have not modelled the
lowest-mass stars at the lowest metallicities because these stars
undergo a helium flash.  The {\sc stars} code is currently unsuitable for
modelling this phase in a star's evolution.  While it is possible to
generate a post-He flash model \citep[see][ for
details]{2007MNRAS.375.1280S}, the method assumes that the envelope
composition is unaffected by the helium flash and this cannot be
guaranteed below a metallicity of $[\rm{Fe}/\rm{H}]<-4.5$
\citep{1990ApJ...349..580F}.

The models are evolved from the pre-main sequence to the TP-AGB with
999 mesh points.  At the beginning of the TP-AGB, the AGB-specific
mesh spacing function of \citet{2004MNRAS.352..984S} is switched on.
It is also at this point that the AGB mixing prescription by the same
authors is applied.  In the initial runs, no convective overshooting
was applied at any phase in the star's evolution.

\section{Results}

We can roughly divide the evolutionary behaviour of these extremely
metal-poor stars into five different regimes, (i) stars with a
flash-driven mixing episode, (ii) stars with CI, (iii) stars with no
third dredge-up, (iv) stars that experience third dredge-up which can
be subdivided depending on whether hot dredge-up or hot bottom burning
or both are present and (v) stars which undergo carbon ignition before
the AGB.  These regimes of behaviour are shown in
Fig.~\ref{fig:ZversusM} in mass and metallicity space.

Stars with an initial mass of 7$\,\rm M_\odot$ or above become
super-AGB (SAGB) when carbon is ignited non-degenerately before
any thermal pulses \citep{2005A&A...433.1037G}.  We do not follow the
evolution of such SAGB stars, owing to the considerable complexity
involved in their computation \citep{2006A&A...448..717S}.  It is
worth noting that a different mixing-length parameter $\alpha$ could
change the critical mass at which a star becomes an SAGB star.  For
example, a 7$\,\rm M_\odot$ model of $Z=10^{-8}$ evolved with
$\alpha=1.925$ does not evolve into a SAGB star.  This is a perfect
example to show that the mixing length parameter is crucial to
determine the evolution of extremely-metal poor stars (and in fact,
any star).  In this work all models are computed with $\alpha=2.0$
and we shall examine the effect of varying it in the future.

Stars with $Z=10^{-4}$ (corresponding to $\rm [Fe/H]\approx-2.3$) behave
similarly to their higher metallicity cousins, with the occurrence of
third dredge up and hot bottom burning.  This suggests that AGB stars
with $Z=10^{-4}$ have enough CNO elements in the hydrogen burning
shell that their evolution is not affected by the low metallicity.
When the metallicity is decreased to $Z=10^{-5}-10^{-6}$, the high
mass AGB stars still behave as they do at higher metallicity, in
that they experience only thermal pulses, third dredge-up and hot
bottom burning.  This is because second dredge-up is more efficient in
higher mass stars and leads to an increased CNO abundance in the
envelope and (more importantly) the hydrogen burning shell (see
Fig.~\ref{fig:Z2dup}).  Therefore, the amount of metals,
particularly the CNO elements in the hydrogen burning shell is one of
the important factors in determining the evolutionary behaviour.

\subsection{Helium-flash-driven mixing and carbon ingestion}

For stars with metallicity at or below $10^{-5}$ and masses at or
below 2$\,\rm M_\odot$ He-FDM can occur
\citep{1990ApJ...349..580F}.  He-FDM occurs when a
convective zone, driven by helium burning, penetrates into
hydrogen-rich regions mixing protons into the carbon-rich parts of the
star.  This can happen either during the helium flash on the red giant
branch or during the TP-AGB phase.  It is the latter case that
concerns us here.  The outer edge of the intershell convective zone
(ICZ), generated by helium burning, extends into layers containing
hydrogen.  Hydrogen is then ingested by the convective zone and is
mixed down to a depth where the temperature is much hotter than the
hydrogen burning shell.  As a result, the hydrogen-burning luminosity
increases substantially.  This leads to the convective zone splitting
into two parts.  The upper convective shell is now driven by hydrogen
burning, while the lower convective shell is still driven by helium
burning.  As more hydrogen is mixed into the hydrogen convective shell
from above, the hydrogen flash develops further.  After the decay of
the flash, the region formerly occupied by the hydrogen-driven
convective shell is carbon- and nitrogen-enhanced because of the
mixing of material from the helium-burning shell.  Then the base of
the convective envelope penetrates into the layers formerly occupied by
the hydrogen burning shell and brings carbon and nitrogen to the
surface.

\begin{figure} 
\includegraphics[width=8.5cm]{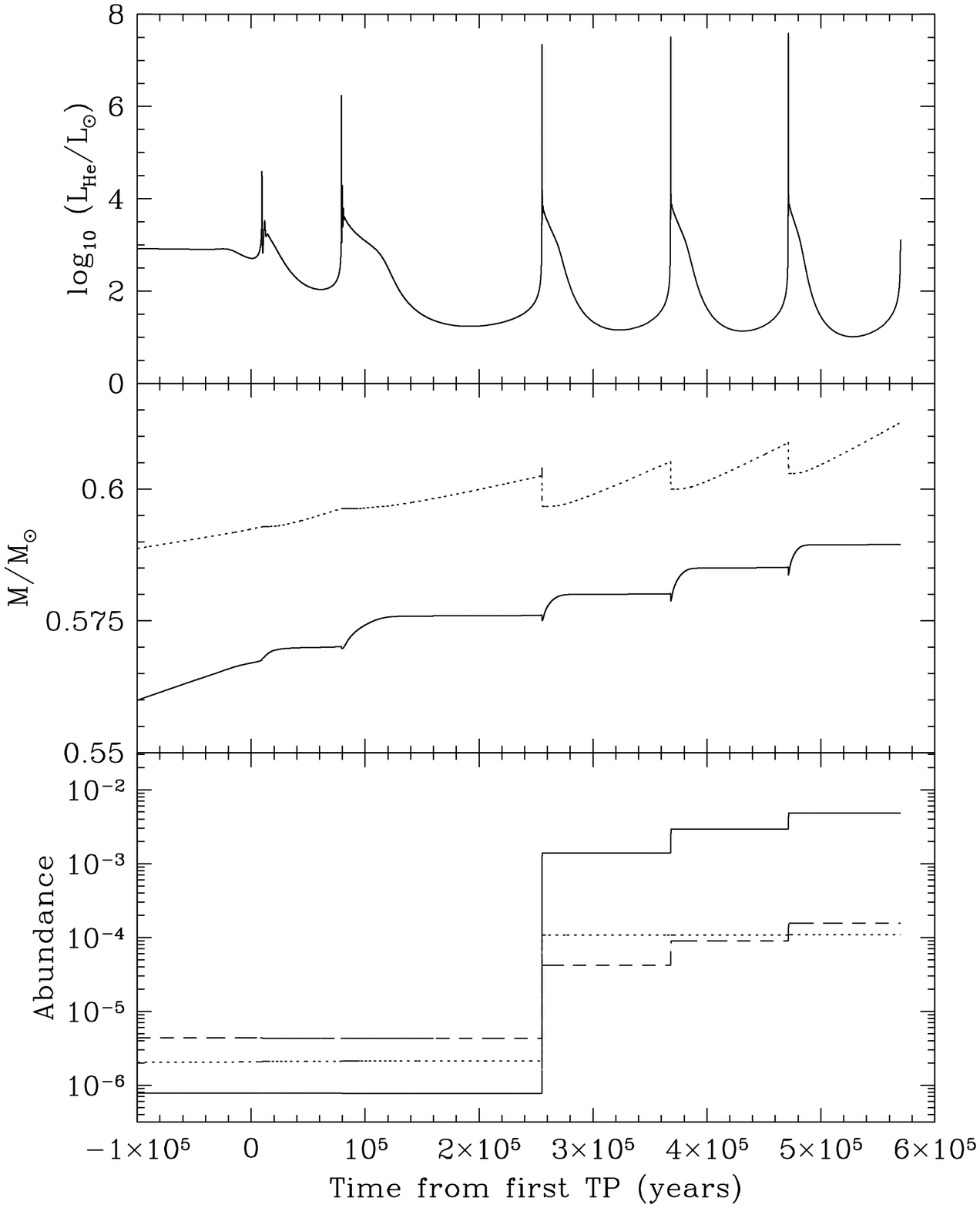}
\caption{Top panel: helium luminosity as a function of time since the
first thermal pulse.  Middle panel: H-exhausted core mass (dotted
line) and He-exhausted core mass (solid line) line as a function of
time.  Bottom panel: abundance by mass fraction of carbon (solid
line), nitrogen (dotted line) and oxygen (dashed line).  These plots
are for the 1.5$\,\rm M_\odot$ model at $Z=10^{-5}$.  The model
undergoes an episode of FDM after about $2.5\times10^{5}\,$yr since
the first thermal pulse, leading to enhancements in the CNO
abundances.  Thereafter, only carbon and oxygen are enhanced.}
\label{fig:m1.5z1e-5}
\end{figure}

\begin{figure}
\includegraphics[width=8.5cm]{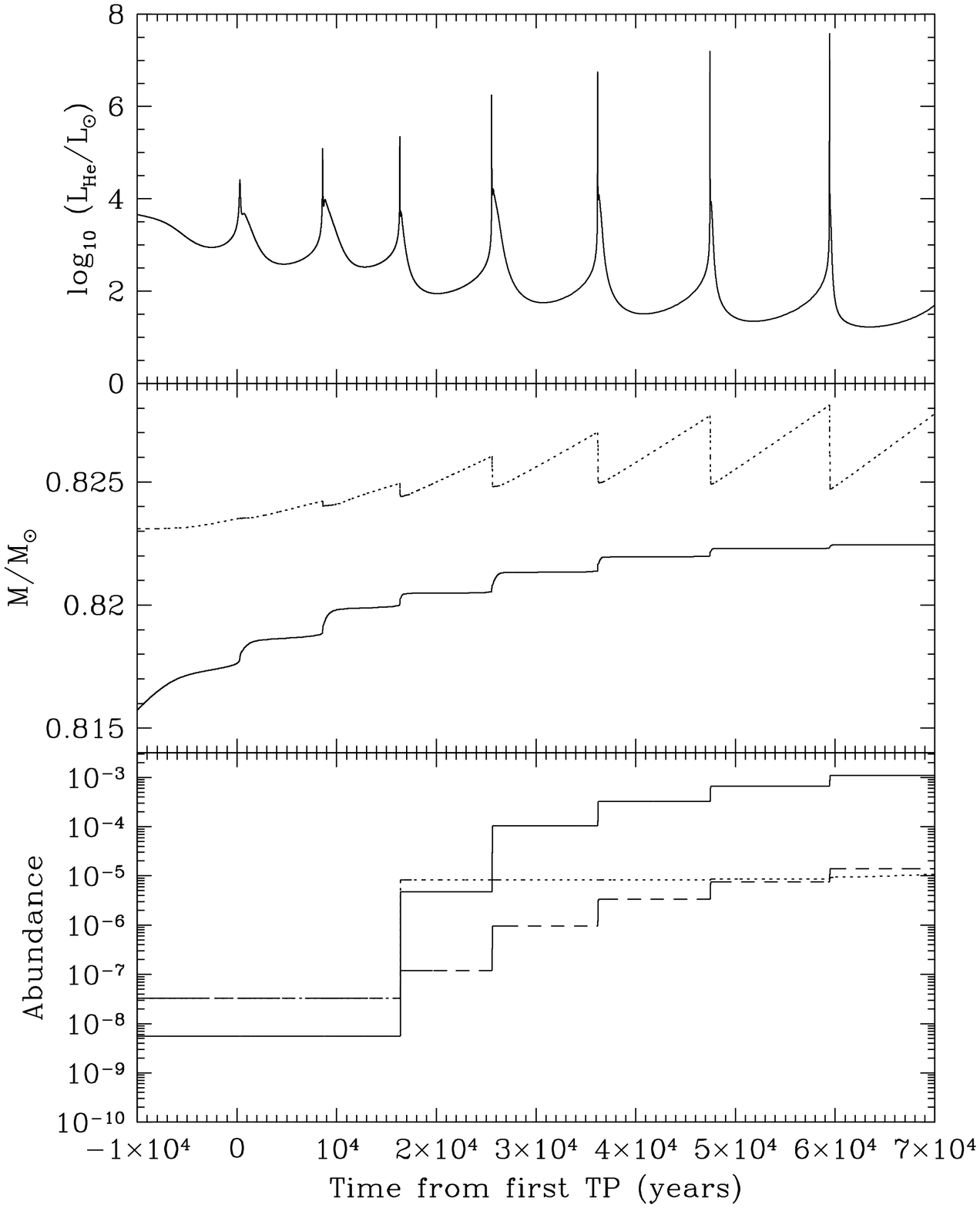}
\caption{Top panel: helium luminosity as a function of time since the
first thermal pulse.  Middle panel: H-exhausted core mass (dotted
line) and He-exhausted core mass (solid line) line as a function of
time.  Bottom panel: abundance by mass fraction of carbon (solid
line), nitrogen (dotted line) and oxygen (dashed line).  These plots
are for the 3$\,\rm M_\odot$ model at $Z=10^{-7}$.  This model
suffers the CI mechanism and the first episode of TDUP
leads to enhancements of C, N and O.}
\label{fig:m3z1e-7}
\end{figure}

\begin{figure} 
\includegraphics[width=7cm, angle=270]{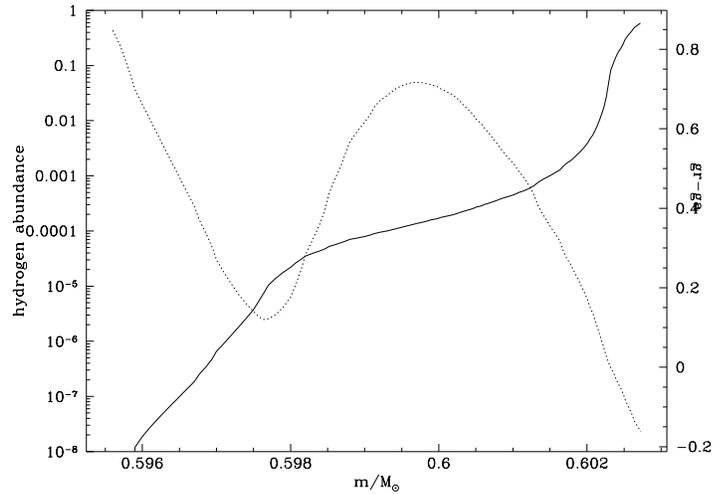}
\caption{FDM hydrogen abundances (solid line) and
$\nabla_{\mathrm{r}}- \nabla_{\mathrm{ad}}$ (dotted line) of our
$1.5\,M_\odot$, $Z=10^{-5}$ model during flash driven mixing.  A region
is convective if $\nabla_{\mathrm{r}}-
\nabla_{\mathrm{ad}}>0$.  The intershell convective zone
reaches the H-burning shell, where the abundance of hydrogen is
around 0.1.  As a result, plenty of protons are mixed down to regions
with higher temperature and carbon abundances.}
\label{fig:FDMs}
\end{figure}

\begin{figure} 
\includegraphics[width=7cm, angle=270]{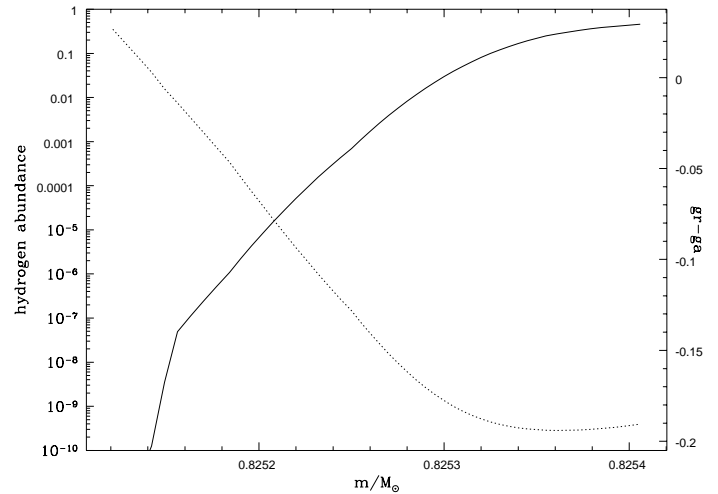}
\caption{CI hydrogen abundances (solid line) and
$\nabla_{\mathrm{r}}- \nabla_{\mathrm{ad}}$ (dotted line) of our
$3\,M_\odot$, $Z=10^{-7}$ model during carbon ingestion.  A region is
convective if $\nabla_{\mathrm{r}}- \nabla_{\mathrm{ad}}>0$.
The intershell convective zone reaches the tail H-burning shell,
where the abundances of hydrogen is around $10^{-7}$.  Compared to
FDM, far fewer protons are mixed, though still enough
to drive a convective zone by the increased hydrogen burning
luminosity.}
\label{fig:CIs}
\end{figure}

Fig.~\ref{fig:1.5FDDM} shows an example of a star undergoing FDM, in
this case a 1.5$\,\rm M_\odot$ star of metallicity $Z=10^{-5}$.  At
around model number 55000, the convective zone, driven by helium
burning, penetrates the hydrogen-burning shell.  This causes the
hydrogen luminosity to peak.  The increased hydrogen luminosity drives
a convective zone which connects with the intershell convective zone.
Material is then mixed between the hydrogen burning shell and the top
of helium burning shells and so there is significant carbon present in
the hydrogen-burning shell.  Later, around model number 60000, the
convective envelope deepens and reaches the region that has been
previously mixed.  In this way, the products of helium burning are
dredged up to the surface during the He-FDM episode.  Nitrogen is also
dredged up because the CNO cycle converts carbon into nitrogen.

Fig.~\ref{fig:m1.5z1e-5} shows that the surface abundances of CNO
elements are enhanced significantly during the He-FDM episode,
corresponding to the third pulse of the $1.5\,M_\odot$, $Z=10^{-5}$
model.  The surface abundance of carbon increases by more than 3
orders of magnitude to the order of $10^{-3}$ because helium burnt
materials are dredged up to the surface.  Nitrogen increases from
about $2\times10^{-6}$ to $10^{-4}$ while oxygen increases by one
order of magnitude to about $4\times10^{-6}$.  It is also worth noting
that after the He-FDM episode, the surface CNO abundances resemble
those of a much higher metallicity star.  As a result of higher CNO
abundances in the hydrogen burning shell, the helium convective zone
cannot any longer reach the hydrogen burning shell in the subsequent
pulses and the star behaves similarly to a higher metallicity star.  Carbon and
oxygen are slowly enhanced during subsequent pulses by third
dredge up just as in higher metallicity stars.

We find that FDM episodes only occur for the lowest stellar masses
(2$\,\rm M_\odot$ and below).  Stars of higher mass do not have
violent enough thermal pulses to drive the ICZ up into the H-depleted
regions.  However they do have pulses that are
violent enough to trigger TDUP.  Once TDUP has raised the CNO
abundance sufficiently, H-burning in the shell proceeds efficiently via
the CNO cycle, the H-shell presents an effective energy barrier to
the ICZ and prevents any FDM.

At $Z=10^{-6}$ there is a change in the evolution of the 3$\,\rm
M_\odot$ and 4$\,\rm M_\odot$ star.  We find that an extra convective zone, located above
the hydrogen burning shell, opens after the peak helium luminosity of
the thermal pulse.  Unlike the FDM mechanism, this convective zone is
not a breakaway region from the intershell convection zone, but is
distinctly separated from it.  This mixing episode is referred to as a
CI by \citet{2001ApJ...554.1159C} and \citet{2002ApJ...570..329S}.  As
described by these authors, carbon from the intershell is injected
into the H-burning shell by this convective zone.  This mechanism also
occurs at $3-6\,\rm M_\odot$, $Z=10^{-7}$ and $3-4\,\rm M_\odot$,
$Z=10^{-8}$ models.  An example of a star that undergoes this process
is shown in Fig.~\ref{fig:m3z1e-7}

We note that there is a distinction between the nature of flash-driven
mixing (FDM) and carbon injection (CI).  The FDM phenomenon involves
protons being mixed down into the intershell region and burning at
very high temperatures.  For the CI phenomenon, mixing occurs when a
convective pocket opens up above the H-shell and its lower edge
penetrates down into C-rich regions.  H-burning takes place at much
lower temperatures than FDM.  In addition, the FDM involves the
ingestion of a small quantity of hydrogen into a carbon-rich region.
Conversely, CI involves the injection of a small amount of carbon into
an H-rich region while the amount of hydrogen mixed is minimal.

In both cases, a convective zone opens up at the top of the hydrogen
burning shell, so when the third dredge-up deepens, both mechanisms
lead to substantial enrichments of nitrogen, though both have only a
single episode of N-enhancement.  Once TDUP has occurred the CNO
abundance is high enough that the star evolves like a higher
metallicity star.  Are these two distinct phenomenon?  Both are
triggered by the ingestion of protons in to the intershell convection
zone \citep[see e.g.][ for the case of the CI]{2002ApJ...570..329S}.
Perhaps it is just the quantity that is ingested that determines which
episode a given star experiences.

\begin{figure*} 
\includegraphics[width=14cm, angle=270]{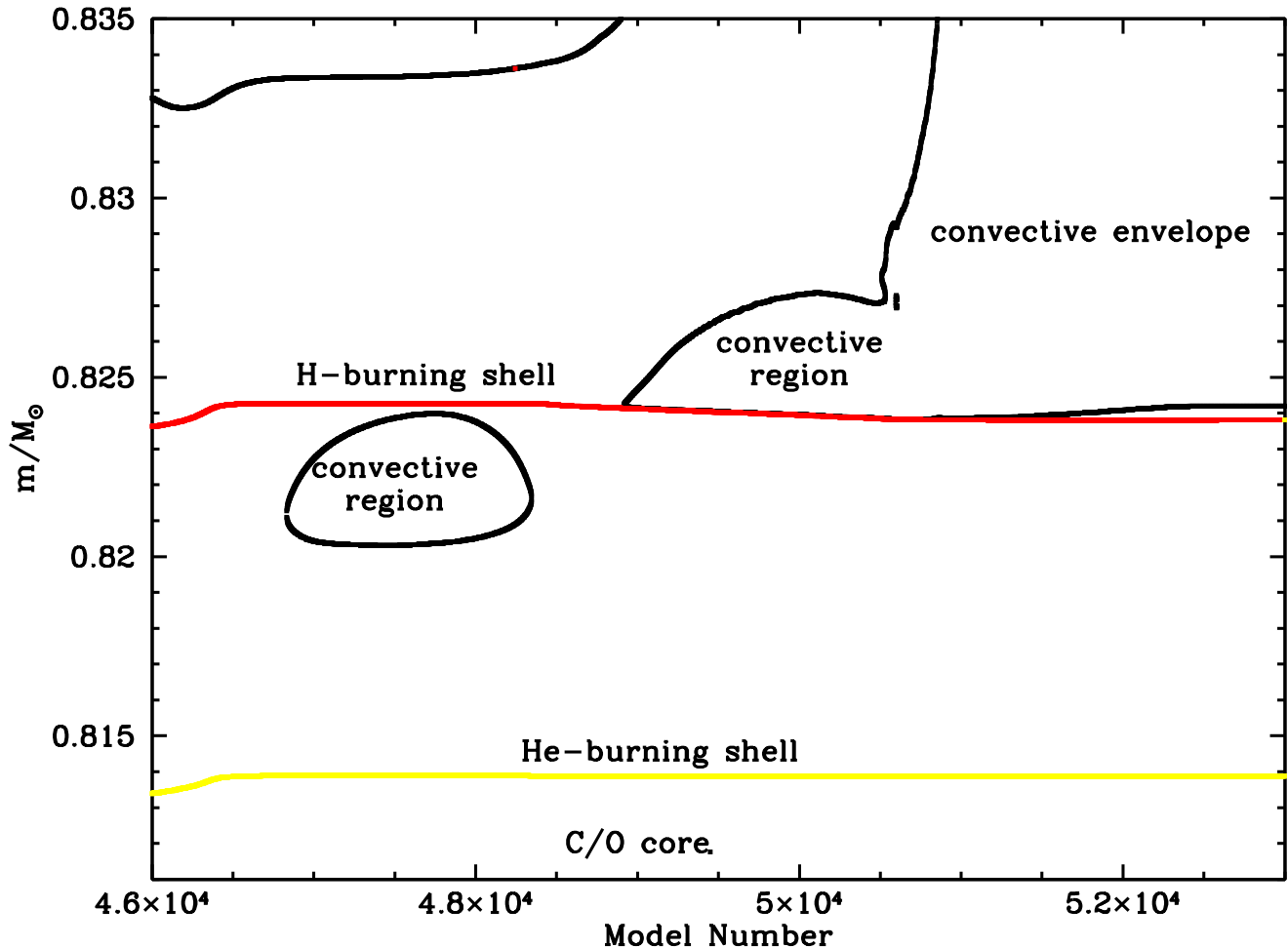}
\caption[Evolution of a$3\,M_\odot$, $Z=10^{-7}$ model during CI]
{Evolution of a $3\,M_\odot$, $Z=10^{-7}$ model during CI.  We can see
the intershell convective zone reaches the tail H-burning shell at
model number around 47000.  Then a convective region driven by
hydrogen burning reaches deep enough to previously mixed region by the
intershell convective zone around model number 50000.  At around model
number 50500, the convective envelope deepens and connect with the
convective zone driven by hydrogen burning.  The third dredge up thus
dredged up carbon and nitrogen indirectly}
\label{fig:3z-7test}
\end{figure*}

\begin{figure}
\includegraphics[width=8.5cm]{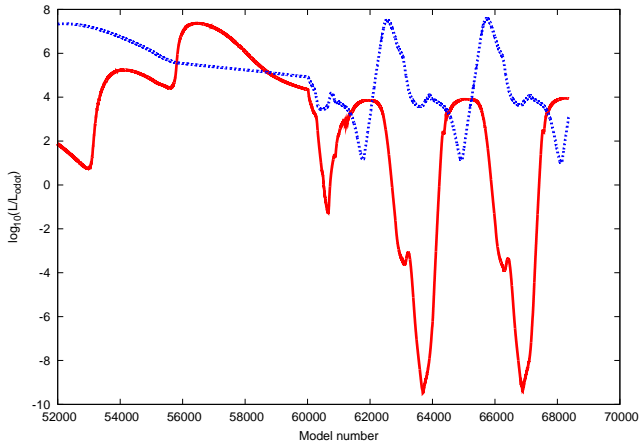}
\caption{Hydrogen luminosity (solid line) and helium luminosity (dotted line)
evolution of a $1.5\,\rm M_\odot$ model at $Z=10^{-5}$.  The hydrogen
luminosity reaches $10^{7}\,{\rm L}_\odot$ during the FDM at around model number
56,000.}
\label{fig:1.5he}
\end{figure}

\begin{figure}
\includegraphics[width=8.5cm]{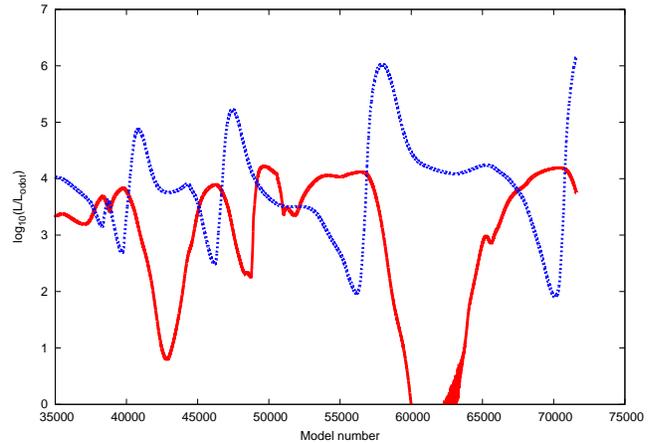}
\caption{Hydrogen luminosity (solid line) and helium luminosity (dotted
line) evolution of a $3\,\rm M_\odot$ model at $Z=10^{-7}$.  The
hydrogen luminosity rises to a second peak of $10^{4}\,{\rm L}_\odot$ (around model
number 48,000) in the second pulse and this does not occur in the
other pulses.  It corresponds to the CI.}
\label{fig:3z-7he}
\end{figure}

We suggest that proton ingestion and He-FDM may have the same origin,
namely the mixing of protons into a carbon rich region but the amount
of hydrogen and carbon mixed determines whether the two convective
zones connect each other and hence whether CI or He-FDM occurs.
Figs~\ref{fig:FDMs} and~\ref{fig:CIs} show how the convective zone
reaches the hydrogen burning shell in the FDM and CI events
respectively.  The convective region penetrates the burning shell and
reaches regions where the hydrogen abundance is as high as 0.1 during
a FDM event, while it only reaches the tail of the burning shell,
where the hydrogen abundance is about $10^{-7}$, in the CI episode.
In both cases, the mixing of protons and carbon causes an increase of
hydrogen luminosity.  In the case of FDM the burning is much more
vigorous so a larger convective zone is driven and the mixing between
the hydrogen burning and helium burning layer is more efficient.

Figs~\ref{fig:1.5FDDM} and~\ref{fig:3z-7test} show the internal
structure around hydrogen burning shell of two stars during FDM and CI
respectively.  In the case of FDM, the vigorous hydrogen burning
immediately drives a convective zone that connects with the intershell
convective zone.  In the case of CI, the burning-driven convection
zone appears later in time and is never connected with intershell
convective zone while it is extant.  FDM mixes material from the
H-burning shell with material from He-burning shell in a single large
convection zone.  For CI the mixing is a two-step process.  As a
result, when third dredge up occurs later, more carbon is dredged up
to the surface in a FDM episode than a CI episode.

In Figs~\ref{fig:1.5he} and~\ref{fig:3z-7he}, we see there is a
increase of hydrogen luminosity after a thermal pulse which correspond
to when FDM and CI occurs.  This is unusual.  The hydrogen luminosity
usually decreases monotonically after a thermal pulse as in the later
pulses.  We can see that the increase of hydrogen luminosity occurs
after the intershell convective zone reaches the hydrogen burning
shell by comparing the model numbers with Figs~\ref{fig:1.5FDDM}
and~\ref{fig:3z-7test}.  The increase of hydrogen luminosity starts at
model number 55580 for the FDM and at model 48672 for the CI.
Compared with Figs~\ref{fig:1.5FDDM} and~\ref{fig:3z-7test}, we can
see that the increases of luminosity just precedes the opening of the
second convection zone around the hydrogen shell.  This is because the
convective zone around hydrogen burning shell is driven by the
increased hydrogen luminosity following the mixing by intershell
convective zone.  The increase of hydrogen luminosity is much higher
in FDM than CI.  The hydrogen luminosity jumps to $10^{7}\,{\rm L}_\odot$ in the
case of FDM and $10^{5}\,{\rm L}_\odot$ in CI.  The peak of the hydrogen
luminosity is unsurprisingly correlated with the number of protons
mixed into the burning region.  The FDM case shows a much stronger
peak because the convective shell penetrates through the hydrogen
burning shell and causes more mixing than following the CI episode.

In terms of the effect on surface abundances, the distinction between
FDM and CI is not so clear cut in some cases.  In fact, a model with
CI but near the FDM-CI boundary can dredge up almost as much carbon as
a model with FDM.  On the other hand, models with CI, but near the
boundary with normal AGB evolution, dredge no more than in normal
third dredge-up.  In some situations, more nitrogen than carbon is
dredged up after CI.  This is because the star also undergoes hot
bottom burning which convert carbon into nitrogen.  In general, the
effect of CI is stronger when metallicity is lower.  This could be
because at lower metallicity, the abundance of carbon at the 
hydrogen burning shell is lower, so the carbon ingested causes a much
larger increase in the hydrogen luminosity, which drives a more
extensive convective zone.

\subsection{Hot dredge-up}

\begin{figure}
\includegraphics[width=8.5cm]{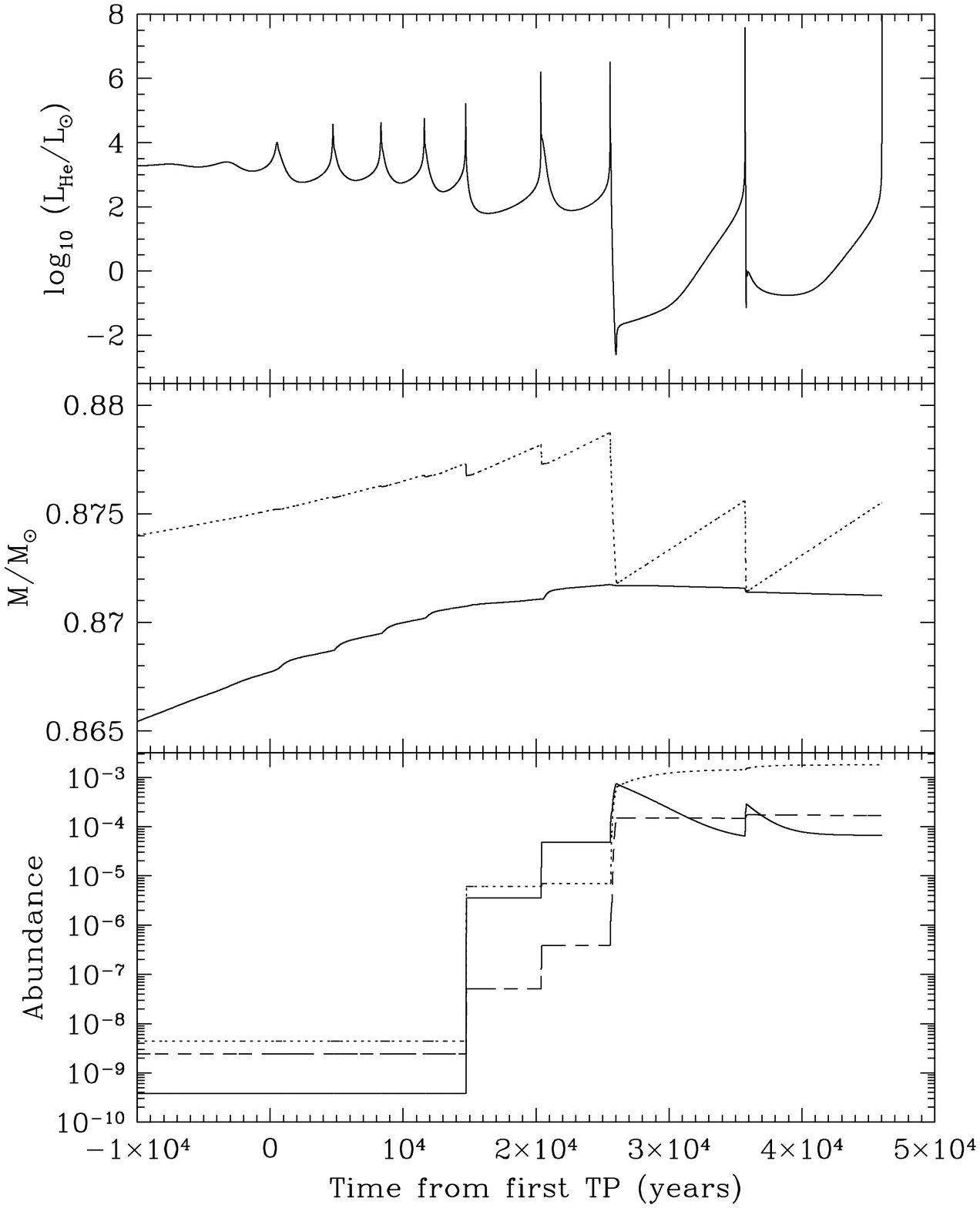}
\caption{Top panel: helium luminosity as a function of time since the
first thermal pulse.  Middle panel: H-exhausted core mass (dotted
line) and He-exhausted core mass (solid line) line as a function of
time.  Bottom panel: abundance by mass fraction of carbon (solid
line), nitrogen (dotted line) and oxygen (dashed line).  These plots
are for our 4$\,\rm M_\odot$ model at $Z=10^{-8}$.  The first episode
of TDUP, while extremely shallow, dramatically enhances the CNO
abundances.  HTDUP sets in after about 25,000\,yr, leading to a
simultaneous increase in carbon and nitrogen before HBB takes over in
the interpulse.}
\label{fig:m4z1e-8}
\end{figure}

In stars of 4$\,\rm M_\odot$ and above, we find some differences in
the dredge-up process.  For the more massive stars, we note the
occurrence of what has been referred to as hot third dredge-up (HTDUP).
This phenomenon was described in detail by
\citet{2004ApJ...605..425H}.  Here, the temperature at the base of the
convective envelope at the point when third dredge-up occurs is so
high that carbon is converted to nitrogen, resulting in a simultaneous
enhancement of both carbon and nitrogen.  HTDUP is an extremely efficient
dredge-up, with the envelope reaching almost to the bottom of the
intershell (or perhaps beyond it, if overshooting is included, see
\citealt{2004ApJ...605..425H} for further details).  The behaviour of
an example of a star that undergoes this process is shown in
Fig.~\ref{fig:m4z1e-8}.  CI occurs at the first episode of third
dredge up in this model and the CNO abundances are drastically
increased.  Then hot third dredge up sets in after two more pulses,
leading to simultaneous increase in carbon and nitrogen.  Hot bottom
burning converts carbon into nitrogen and nitrogen becomes more
abundant than carbon by more than one order of magnitude.

\subsection{A metallicity threshold for dredge-up?}
As we move to the lowest metallicity, we find that TDUP ceases
altogether for some models.  The 5$\,\rm M_\odot$ model, with
$Z=10^{-8}$ has very weak thermal pulses and these are not strong
enough to bring the convective envelope into contact with the C-rich
intershell.  On the other hand, the 6$\,\rm M_\odot$ model, with
$Z=10^{-8}$ does have a third dredge up.  The CNO content of the
6$\,\rm M_\odot$ in the hydrogen burning shell is higher than the
5$\,\rm M_\odot$ because of deeper second dredge up.  This seems to
confirm the strength of the pulses are dependent on the CNO abundances
in the hydrogen burning shell, as suggested by
\citet{2008MNRAS.385..301L}.

The issue of the efficiency of third dredge-up is a perennial problem
for modellers of AGB stars.  Different stellar evolution codes
give very different predictions for the efficiency of TDUP.  Some of
the reason for this uncertainty is due to the mathematical treatment
of the boundaries between radiative and convective regions.  Whether
one includes convective overshooting at the boundary may also affect
the depth of dredge-up.  For example, \shortcite{2001ApJ...554.1159C}
find that third dredge up does occur in their zero-metallicity
intermediate mass AGB stars.  \shortcite{2001ApJ...554.1159C} treated
the convective boundaries according to the prescription of
\shortcite{Herwig}, who use a mixing scheme so efficient that the
composition discontinuity between the two burning shells is smoothed
out.  This seems to indicate the efficiency of third dredge-up depends
on the treatment of convection and the inclusion of extra-mixing
mechanisms such as convective overshooting.  However,
\shortcite{Gilpons} find that the total amount of mass dredged-up is
very small even when overshooting is included.  The {\sc stars} code
employs an arithmetic mean when computing the mixing coefficient at
the convective boundaries.  While this means we are mixing in material
that lies outside the formal Schwarzschild boundary, it is a far
smaller degree of overshooting than applied by other authors but a
degree that leads to a stable composition and convection profile.  We
discuss the effects of overshooting in more detail in
Section~\ref{sec:OS}.  We find that the inclusion of convective
overshooting in our model is enough to cause third dredge
up to occur at $Z=10^{-8}$.

See \citet{2008MNRAS.385..301L} for a more detailed discussion and
description of the absence of third dredge-up.  We notice that the
5$\,\rm M_\odot$, $Z=10^{-8}$ star does not show third dredge-up. 
Also thermal pulses are growing weaker and weaker
and eventually stop after 560 pulses in $5.5 \times 10^{5}yr$. Then the star enters a quiescent burning phase which lasts until carbon ignites degenerately at the centre when the CO core mass is 1.36$\,\rm M_\odot$
This agrees with the conclusion that supernovae Type
1.5 are likely to be the fate of these stars because of the faster
core growth rate when thermal pulses cease
\citep{2008MNRAS.385..301L}.

\subsection{2nd dredge up}

Fig.~\ref{fig:Z2dup} shows the abundance of the CNO elements at the
surface of each star at the beginning of the TP-AGB.  We note that
second dredge-up can lead to substantial enhancement of the CNO
abundance.  Of particular note is the 5$\,\rm M_\odot$, $Z=10^{-6}$
model, which has a surface CNO abundance that is about ten times
higher than its initial metallicity.  This
enhancement in metallicity is sufficient to allow the star to
experience a relatively normal TP-AGB.  More CNO strengthens the
entropy barrier and decreases the H-tail from the H-burning shell.
Note that at the same metallicity, the 3$\,\rm M_\odot$ star
experiences a CI phenomenon.  Also, for the same mass, as we move to
lower metallicity, the behaviour changes suddenly as we drop the initial
metallicity by just a factor of ten.

\subsection{Zero-metallicity models}

For stars at zero metallicity, the behaviour is similar to stars with
$Z=10^{-8}$.  In the 2$\,\rm M_\odot$ model, there is FDM at the
first thermal pulse.  For the 3$\,\rm M_\odot$ model, the evolution
is much more complicated.  During the first pulse there is a CI phase
but it is rather weak and does not increase the surface metallicity by
much.  There is then FDM at the second thermal pulse.  The 4$\,\rm
M_\odot$ model encounters CI in the first pulse.  The thermal pulses
of the 5$\,\rm M_\odot$ and 6$\,\rm M_\odot$ model are not strong
enough to cause any third dredge up and probably end their lives as
type 1.5 supernovae \citep{2008MNRAS.385..301L}.

\subsection{Comparison to models from literature}

\citet{2008A&A...490..769C} have published a grid of models
for $M/{\rm M}_\odot = 0.85$, 1.0, 2.0, 3.0 and the metallicity
$[\rm{Fe}/\rm{H}] = -6.5$, -5.45, -4.0, -3.0 and $Z=0$.  Our results
for 2$\,\rm M_\odot$ models agree well with their models.  They find
FDM (referred as ``Dual Shell Flash'' by them) at metallicity
$[\rm{Fe}/\rm{H}] = -4.0$ or below but not at metallicity
$[\rm{Fe}/\rm{H}]=-3.0$.  This agrees with our result that the
critical metallicity for FDM to occur is between $Z=10^{-6}$
to~$10^{-5}$, corresponding to $[\rm{Fe}/\rm{H}]$ between -4.3
and~-3.3.  For their 3$\,\rm M_\odot$ models FDM does not occur at
$[\rm{Fe}/\rm{H}] = -4.0$ or above but does when $[\rm{Fe}/\rm{H}] =
-5.45$ and below.  Our result is somewhat different.  We only find the
milder CI episode instead of FDM in all our 3$\,\rm M_\odot$ models
except when $Z=0$.  CI begins to occur between $Z=10^{-6}$
and~$10^{-5}$ ($[\rm{F e}/\rm{H}]= -4.3 - -3.3$) and this agrees with
their metallicity range for FDM at 3$\,\rm M_\odot$.  For 1$\,\rm
M_\odot$ models we find the critical metallicity for flash driven
mixing to be between $Z=10^{-5}$ and~$Z=3\times10^{-5}$
($[\rm{Fe}/\rm{H}]= -3.3 - -2.8$).  This agrees with the fact that FDM
already occurs in their model with metallicity $[\rm{Fe}/\rm{H}] =
-3.0$.

One significant difference between the two sets of models is the issue
of hot bottom burning.  Hot bottom burning does not occur for any of
our models with initial mass 3$\,\rm M_\odot$ or below, while their
2$\,\rm M_\odot$ and 3$\,\rm M_\odot$ models have strong hot bottom
burning.  Therefore, nitrogen is more abundant in their 2$\,\rm
M_\odot$ and 3$\,\rm M_\odot$ models, while carbon is more abundant
in our models.  Hence, it is easier to form CEMP stars with
$[\rm{C}/\rm{N}] \approx 1$ from our models.

\citet{Komiya} give an up-to-date version of the dependence of FDM
and the third dredge up on the initial stellar mass and metallicity,
formulated by \citet{2000ApJ...529L..25F}.  Our models show a few
differences to their picture.  Their critical metallicity for FDM to
occur is $[\rm{Fe}/\rm{H}] = -2.5$.  This is higher than the critical
metallicity found in our grid of models ($[\rm{Fe}/\rm{H}] = -3.3$
to~$-2.8$).  Moreover, in their picture, the critical metallicity for
FDM to occur is the same for initial masses between 0.8$\,\rm M_\odot$
to~3.5$\,\rm M_\odot$.  This is not the case in our models
(Fig.~\ref{fig:ZversusM}) where the critical metallicity for FDM to
occur in 2$\,\rm M_\odot$ stars is lower than that of 1$\,\rm M_\odot$
and 1.5$\,\rm M_\odot$ stars.  Our mass range for FDM is also smaller
than their picture.

Another important difference is the nitrogen surface abundances.
Their picture shows that extremely metal-poor stars between 3.5 $\,\rm
M_\odot$ to 5$\,\rm M_\odot$ could be carbon rich but nitrogen normal.
However, most of our $4-6\,\rm M_\odot$ models are nitrogen rich
due to hot dredge-up.  We agree that there are metal-poor stars that
have no third dredge up but the mass range from our models is much
smaller than theirs.

\section{The effect of overshooting}
\label{sec:OS}

The above behaviour raises an interesting question: what is the effect
of the inclusion of convective overshooting on the behaviour of the
models? To examine this, we have constructed a grid of models with
convective overshooting.  We use the prescription of
\citet{1997MNRAS.285..696S} and their overshooting parameter,
$\delta_\mathrm{ov}=0.12$.  This overshooting is applied up to the
beginning of the TP-AGB in all models made.

\begin{figure*}
\includegraphics[width=14cm, angle=0]{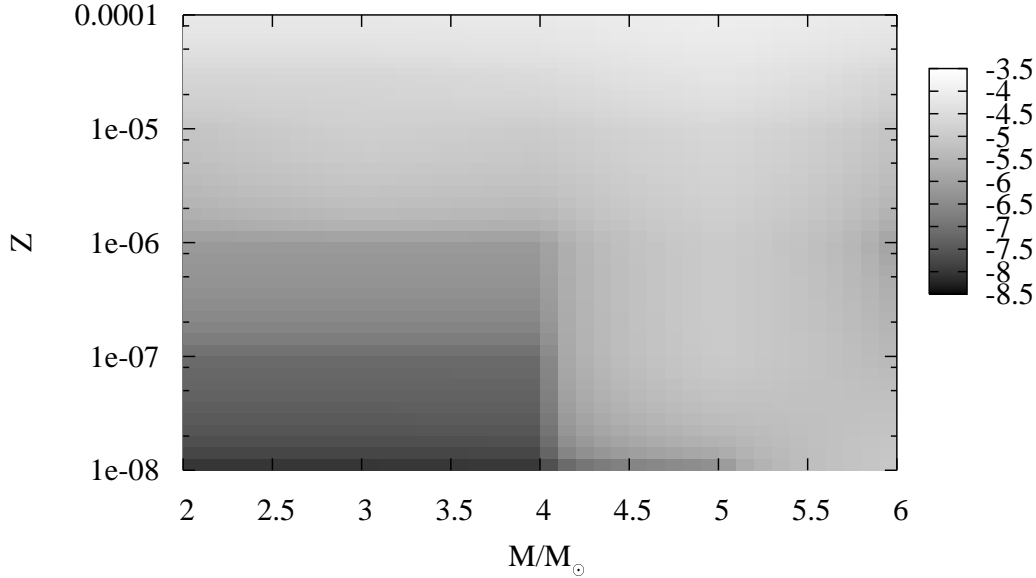}
\caption{Plot of the logarithm of the abundance of CNO elements by
mass at the beginning of the TP-AGB.  As is the case of the models
without overshooting, the action of second dredge-up substantially
raises the abundance in the case of the intermediate mass stars.  The
additional overshooting makes the effects more pronounced here.}
\label{fig:Z2dupOs}
\end{figure*}

Fig.~\ref{fig:Z2dupOs} shows the abundance of the CNO elements at
the surface of each star at the beginning of the TP-AGB.  As with the
non-overshooting case, we find that second dredge-up can lead to
substantial enhancement of the CNO abundance.  The effect is even more
substantial with the inclusion of overshooting.  This is what we would
expect to find.  We also note that we get a substantial elevation of
the CNO abundance at lower initial masses.  With overshooting our
$5\,\rm M_\odot$, $Z=10^{-8}$ model is significantly enhanced in CNO
elements.  The CNO abundances of $4-6\,\rm M_\odot$,
$Z=10^{-7}-10^{-8}$ models with overshooting prescription is higher
than models without overshooting by about
one order of magnitude.

Unfortunately we are not able to evolve some models along the whole
TP-AGB with convective overshooting, owing to as yet unresolved
numerical difficulties during hot dredge-up for some stars.  These
models fail to converge at the onset of hot dredge-up if overshooting
is applied.  We therefore evolve without the overshooting just before
the start of hot dredge-up.  Fortunately we can evolve some models,
such as the $4\,\rm M_\odot$, $Z=10^{-5}$, through a significant
number of thermal pulses with overshooting and, based on these models,
we believe the inclusion of overshooting during the TP-AGB phase does
not significantly affect the subsequent evolutionary behaviour.

\begin{figure*}
\includegraphics[width=14cm, angle=270]{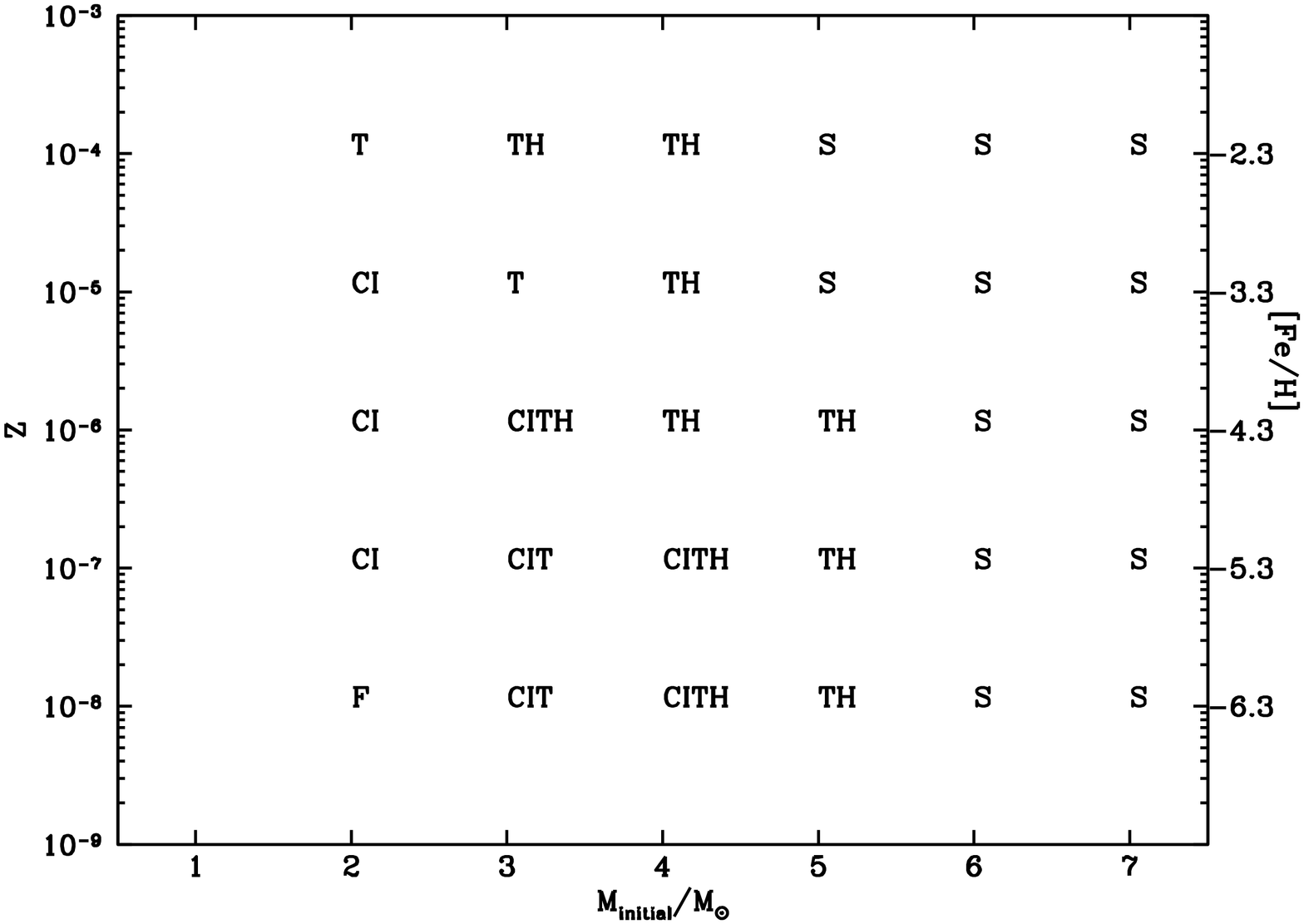}
\caption{The AGB behaviour of the models as a function of mass and
metallicity with overshooting.  The symbols are T -- third dredge-up,
H -- hot third dredge-up, F -- FDM, CI -- CI, X -- no TDUP and S -- CI
before the AGB (presumably leading to an S-AGB star).}
\label{fig:ZversusM2}
\end{figure*}

Models with overshooting have higher core masses when they enter the
TP-AGB phase because of the more extended convective cores during core
He-burning.  Also, the amount of CNO elements dredged up is higher
when overshooting is included.  As a result, a model with a given mass
and metallicity with overshooting behave more like a model with higher
mass without overshooting.  Comparing the evolutionary behaviour with
overshooting and without overshooting, we see that the major
effect is to shift the boundary of different regimes
of behaviour by mass but not metallicity (Fig.~\ref{fig:ZversusM2}).

We do not find any flash driven mixing or carbon ingestion in stars
with $Z=10^{-4}$ as in the case of models without overshooting.  At
$Z=10^{-5}$ the overshooting models of 2$\,\rm M_\odot$ show the
occurrence of a relatively weak CI, which is absent in the models
without overshooting.  The overshooting prescription of {\sc stars} code
mixes regions where $\nabla_{\mathrm{r}}- \nabla_{\mathrm{ad}}
>-0.12$.  Therefore, regions that were close to instability can become
convective.  As a result, a small convective region is opened up and
carbon is mixed into proton-rich region.

Unlike models without overshooting, models from $Z=10^{-6}$ to
$Z=10^{-7}$ do not show any signature of FDM.  It is possible that the
maximum mass for a model with overshooting to have FDM is just below
2$\,\rm M_\odot$ for metallicities between $10^{-6}$ and $10^{-7}$
because a strong CI episode is found in the 2$\,\rm M_\odot$ models at
these two metallicities.  Flash driven mixing does occur in our
2$\,\rm M_\odot$ model with metallicity at $10^{-8}$.

\begin{table*}
\begin{center}
\begin{tabular}[h] {c c c c c c}
\hline
Mass/$M_\odot$  & metallicity & ${}^{12}\rm{C}$ & ${}^{14}\rm{N}$& ${}^{16}\rm{O}$ & $\lambda$\\
\hline
3.0 & $10^{-4}$ &$ 2.4 \times 10^{-4}$ & $2.6 \times 10^{-3}$     & $3.0 \times  10^{-4}$ &$3.7$\\
3.0 & $10^{-6}$ &$ 2.6 \times 10^{-4}$ & $3.0 \times 10^{-3}$     & $2.8 \times  10^{-4}$ &$3.3$\\
\hline
4.0 & $10^{-4}$ &$ 7.1 \times 10^{-4}$ & $8.6 \times 10^{-4}$     & $8.1 \times  10^{-5}$ &$5.7$\\
4.0 & $10^{-5}$ &$ 3.7 \times 10^{-5}$ & $8.7 \times 10^{-4}$     & $7.4 \times  10^{-5}$ &$7.4$\\
4.0 & $10^{-6}$ &$ 3.8 \times 10^{-5}$ & $9.3 \times 10^{-4}$     & $9.8 \times  10^{-5}$ &$5.3$\\
4.0 & $10^{-7}$ &$ 3.9 \times 10^{-5}$ & $1.0 \times 10^{-3}$     & $1.0 \times  10^{-4}$ &$6.1$\\
4.0 & $10^{-8}$ &$ 1.8 \times 10^{-5}$ & $4.3 \times 10^{-4}$     & $1.4 \times  10^{-5}$ &$4.1$\\
\hline
5.0 & $10^{-6}$ &$ 1.6 \times 10^{-5}$ & $2.4 \times 10^{-4}$     & $9.0 \times  10^{-6}$ &$4.3$\\
5.0 & $10^{-7}$ &$ 1.2 \times 10^{-5}$ & $2.8 \times 10^{-4}$     & $1.0 \times  10^{-5}$ &$4.5$\\
5.0 & $10^{-8}$ &$ 2.7 \times 10^{-5}$ & $6.2 \times 10^{-4}$     & $5.8 \times  10^{-5}$ & $2.3$ \\

\hline
\end{tabular}
\caption{Surface abundances by mass fraction after the first hot
dredge-up episode and dredge-up efficiency $\lambda$ of the hot dredge-up episode for models with overshooting.}
\end{center}
\end{table*}

There is a general trend that the upper mass limit for CI increases as
metallicity decreases.  At $Z=10^{-5}$ only the 2$\,\rm M_\odot$ model
shows a weak CI episode.  CI occurs in both the 2$\,\rm M_\odot$ and
3$\,\rm M_\odot$ models at $Z=10^{-6}$ and in the 2$\,\rm M_\odot$,
3$\,\rm M_\odot$ and 4$\,\rm M_\odot$ models at $Z=10^{-7}$.  At
$Z=10^{-8}$, the 3$\,\rm M_\odot$ and 4$\,\rm M_\odot$ models still
encounter CI but the 2$\,\rm M_\odot$ model has a flash-driven mixing
episode.

It is worth noting that models of $6-7\,\rm M_\odot$ ignite carbon
before any thermal pulses and become super-AGB stars.  The 5$\,\rm
M_\odot$ models at $Z=10^{-4}$ and $Z=10^{-5}$ also enter the SAGB
phase before any thermal pulses and we do not model the evolution of
these stars.  This is in agreement with the general effect of
overshooting which causes models to behave like higher mass models.

We also find hot dredge-up in most of the $3-5\,\rm M_\odot$ models,
compared to the mass range of $4-6\,\rm M_\odot$ in the
non-overshooting models.  All of the $4\,\rm M_\odot$ models encounter
hot dredge up.  While the two 5$\,\rm M_\odot$ models with $Z=10^{-4}$
and $10^{-5}$ enter the SAGB branch, the other three 5$\,\rm M_\odot$
models with lower metallicities also encounter third dredge up.  For
the five 3$\,\rm M_\odot$ models, we only find hot dredge up in the
$Z=10^{-4}$, $Z=10^{-6}$ models.  Nevertheless, we believe hot
dredge up should occur in all 3$\,\rm M_\odot$ stars with metallicity
ranging from $10^{-4}$ to $10^{-8}$.  We did not find the occurrence
in some of the 3$\,\rm M_\odot$ models because of the earlier
break-down of the evolutionary code owing to unsolved numerical
problems.  

Hot dredge up occurs at around the 5th pulse at
metallicity of $10^{-4}$ and $10^{-5}$.  At low metallicity hot dredge
up tends to occur in later pulses.  In the $5\,\rm M_\odot$, $Z =
10^{-8}$ model it occurs at the 12th pulse.  The dredge up efficiency
$\lambda$, defined as the mass dredged up divided by the core mass
growth during interpulse phase, is highest for $4\,\rm M_\odot$ models
in general (see table 1).  The dredge up efficiency seems to remain
relatively constant between $10^{-4}$ and $10^{-7}$ but is lower at
$Z=10^{-8}$.  For 5$\,\rm M_\odot$ models the dredge up efficiency is
also the same for the $Z=10^{-6}$ and $Z=10^{-7}$ models but is lower
in $Z=10^{-8}$.  For the two 3$\,\rm M_\odot$ models, which show hot
dredge up, $\lambda$ is also similar.  In most cases, the code fails
to converge at the next pulse after the hot dredge up but in a few,
such as the 4$\,\rm M_\odot$, $Z=10^{-6}$ model, we manage to evolve
for two pulses with hot dredge up.  The dredge up efficiency is much
lower in the 2nd pulse, 1.5 compared to 5.3 in the previous pulse.

Also of interest is that the surface abundances after hot dredge
do not depend much on metallicity.  For 4$\,\rm M_\odot$
models, the surface abundances of carbon are almost constant at $3.8
\times 10^{-5}$ for $Z=10^{-5}, 10^{-6}, 10^{-7}$ models (see Table
1).  For nitrogen and oxygen, the surface abundances also hardly vary.
At $Z=10^{-4}$ model there is more carbon enhancement than for the 4$\,\rm
M_\odot$ models at lower metallicities while the nitrogen and oxygen
surface abundances are roughly the same for models with different
metallicities.  At $Z=10^{-8}$ the surface abundances of carbon and
nitrogen are about half those at higher metallicities while oxygen is
down by nearly a factor of ten.  The CNO abundances for our $5\,\rm
M_\odot$, $Z=10^{-6}$ and $Z=10^{-7}$ stars are almost the same while
the $Z=10^{-8}$ model enriches its surface with more CNO elements by
several times.  The surface CNO abundances are also very close in the
3$\,\rm M_\odot$, $Z=10^{-4}$ and $Z=10^{-6}$ models.  This indicates
that hot third dredge up wipes out any differences in the initial
metallicity of the stars and that subsequently the surface abundances
mainly depend on the initial mass.

Amongst our overshooting models we do not find any that do not
experience third dredge-up.  One possible reason is that the second
dredge up manages to increase the surface metallicity, in particular
the CNO elements in the hydrogen shell, above the threshold at which
the hydrogen burning shell can be fully sustained by the CNO cycle
\citep{2001ApJ...554.1159C}.

\section{CEMP stars}

From our models, we can see that there are a few phenomena that
can affect the composition of the material dredged up, namely
flash-driven mixing, CI and hot dredge-up.  If the material from the
envelopes of these stars is accreted on to a binary companion
it becomes carbon enhanced.  In this section,
we compare the surface abundances of our AGB star models results from
our code to the observed abundances of different types of CEMP stars
to check what types of CEMP stars can be formed by the binary transfer
scenario.  Objects used in this section are picked from the Stellar
Abundances of Galactic Archaeology (SAGA) Internet database
\citep{2008PASJ...60.1159S} and are chosen because their observed
abundances fit well the abundances of our models.

\citet{2008MNRAS.389.1828S} have discussed the formation of CEMP
stars by mass transfer from AGB stars in the mass range 1$\,\rm
M_\odot$ to 3.5$\,\rm M_\odot$ with the effect of thermohaline mixing
in the secondary.  One of the problems they note with the abundances
from the AGB models is that they do not predict substantial
enhancements of carbon and nitrogen at the same time except in a very
narrow mass range.  This problem is not solved by the evolution of
higher mass AGB models in this work.  Although carbon can be brought
to the surface in third dredge up in these higher mass models, the
presence of hot bottom burning destroys most of it.  For example, in
the $5\,\rm M_\odot$, $Z=10^{-4}$ star without overshooting,
$[\rm{C}/\rm{Fe}]$ is roughly 1 and $[\rm{N}/\rm{Fe}]$ is greater than
3 after hot third-dredge up and hot bottom burning.  In the
overshooting model, the surface abundances of carbon and nitrogen of
the 4$\,\rm M_\odot$ model are roughly equal.  In this model,
$[\rm{C}/\rm{Fe}]$ is equal to 1 .6 and $[\rm{N}/\rm{Fe}]$ is equal to
2.2 when $[\rm{Fe}/\rm{H}]$ is about $-2.4$.  This could possibly fit
the observed abundances of CEMP stars with strong nitrogen
enhancement, such as HE~1031-0020.  The $3\,\rm M_\odot$, $Z=10^{-4}$
model gives a surface abundances of $[\rm{C}/\rm{Fe}]=1.1$ and
$[\rm{N}/\rm{Fe}]=2.7$ and may fit star such as HE~0400-2030.
However, even with overshooting, the mass range in which we predict
substantial enhancement of carbon and nitrogen is indeed narrow.

At a metallicity of $Z=10^{-4}$, we do not find any flash-driven
mixing or carbon ingestion in any of our models.  When we move down to
the metallicity of $Z=10^{-5}$, an AGB star could have flash-driven
mixing or carbon ingestion or hot dredge-up depending on its mass.
More importantly, enhancement of carbon, nitrogen and other elements
differs for these three mechanisms.  For stars with FDM, the violent
nature of the flash causes a very efficient mixing from the helium
burning region to the hydrogen burning region.  Therefore, when the
convective envelope dredges into the region that is previously mixed
by the FDM, significant amounts of the CNO elements are dredged to the
surface.  Carbon can be enhanced by more than 1,000 times
(Fig.~\ref{fig:1.5FDDM}).  Nitrogen and oxygen are not enhanced as
much and the C/N ratio approaches 100.  For example, in the star of
1.5$\,\rm M_\odot$ with $Z=10^{-5}$ subsequent thermal pulses further
increase the carbon and oxygen surface abundance but not nitrogen
because the star is not hot enough for hot bottom burning.  As a
result, the $[\rm{C}/\rm{N}]$ ratio can reach 1.5 or higher.  Also,
the $[\rm{C}/\rm{Fe}]$ could be as high as~3.  This could explain the
formation of HE~1430-0919 which has $[\rm{C}/\rm{Fe}] = 2.7$,
$[\rm{N}/\rm{Fe}] = 1.6$ and $[\rm{Fe}/\rm{H}] =-3.01$.

For stars with a CI episode, we find that there is a range of carbon
and nitrogen enhancement dependent on the mass of the stars.  It is
possible for stars with CI to achieve carbon and nitrogen enhancement
five times less than the stars with FDM with a similar
$[\rm{C}/\rm{N}]$ ratio.  In some cases, such as the 3$\,\rm M_\odot$,
$Z=10^{-5}$ model with overshooting, the amount of carbon dredged up
is lower and $[\rm{C}/\rm{Fe}]$ only reaches~2.0 and
$[\rm{N}/\rm{Fe}]$ only~0.9.  This could well explain the formation of
stars such as HE~2330-0555 and HE~0017+0055.  On the other hand, it is
possible that CI could cause the nitrogen abundance to be similar to
the carbon abundance as in the case of 3$\,\rm M_\odot$, $Z=10^{-7}$
without overshooting after the enhancement by CI in the third pulse
(see Fig.~\ref{fig:m3z1e-7}).  Whether CI preferentially enhances
carbon or nitrogen depends on how efficiently the star converts carbon
into nitrogen and hence depends on the mass and metallicity of the
stars.  It has to be pointed out that, in the case of 3$\,\rm
M_\odot$, $Z=10^{-7}$ without overshooting, the subsequent third
dredge up by later pulses eventually causes the surface carbon
abundance to be higher than the surface nitrogen abundance.

For higher-mass AGB stars at this metallicity, the hot dredge up could
lead to simultaneous enhancement of carbon and nitrogen.  However,
while a significant amount of carbon is dredged up, most of it is
converted to nitrogen by hot bottom burning, leading to a strong
nitrogen enhancement.  For example the 4$\,\rm M_\odot$, $Z=10^{-5}$
stars have surface abundances of $[\rm{C}/\rm{Fe}] = 1.3$,
$[\rm{N}/\rm{Fe}] = 3.1$ and $[\rm{Fe}/\rm{H}] =-3.3$.  Such
nitrogen-rich stars have been searched for but there is an apparent
dearth of them \citep{2007ApJ...658.1203J}.  As the surface abundances
of our models after hot dredge-up does not depend on metallicity,
$[\rm{N}/\rm{Fe}]$ increases with decreasing metallicity and stars
with lower metallicity should have even stronger nitrogen enhancements
if metal-poor AGB stars with hot dredge-up do exist.

As we go down in metallicity to about $10^{-7}$, the second dredge up
of high-mass AGB stars can reach deep enough to penetrate into regions
where helium has been burned.  Even though third dredge-up is weak or
non-existent in such stars, carbon is already enhanced before entering
the TP-AGB phase.  For example, in the 5$\,\rm M_\odot$, $Z=10^{-8}$
model with overshooting, just after second dredge-up, carbon is
enhanced by a factor of $10^3$ while the $\rm{C}/\rm{N}$ ratio is
around 10.  This could potentially be an important way of forming CEMP
stars with extremely low-metallicity, such as HE 0107-5240
\citep{2007MNRAS.378..563L}.  At the moment, we only observe two stars
with an iron abundance of less than $[\rm{Fe}/\rm{H}]  < -5$ 
(corresponding to $Z=10^{-7}$ if all the abundances were
solar-scaled and the enhancement happened after these stars formed),
HE 0107-5240 with $[\rm{Fe}/\rm{H}] = -5.3$ \citep{ChristliebN} and HE 1317-2326
with$[\rm{Fe}/\rm{H}] = -5.45$\citep{Frebel}.  However, this scenario
cannot explain HE 1317-2326, because of the similar observed
enhancement of carbon and nitrogen.  In fact, none of our models at
such a low metallicity manage to produce a surface abundance with that
C/N pattern.

At such low metallicity, the carbon enhancement by high-mass AGB stars
through second dredge up can be as significant as that of low-mass
stars which go through FDM or CI.  At metallicity of $10^{-5}$ and
$10^{-6}$, the low-mass AGB stars have stronger carbon enhancement
than high-mass AGB stars because FDM or CI leads to more efficient
dredge up of carbon than third dredge-up alone.

In conclusion, it is possible to explain CEMP stars with strong carbon
enhancement and weaker nitrogen enhancement.  They are most likely to
be formed from low-mass AGB stars.  At $Z=10^{-4}$ these stars have
third dredge up without hot bottom burning, so carbon is strongly
enhanced.  With metallicity around $Z=10^{-5}$, both FDM and CI lead
to significant carbon enhancement.  It is harder to explain stars that
have significant enhancement of both carbon and nitrogen.  While a
narrow mass range of stars at $Z=10^{-4}$ may be able to enrich carbon
and nitrogen significantly through hot dredge-up, most models with hot
dredge-up destroy too much carbon and enrich with too much nitrogen to
fit the current observations.  At a metallicity of $10^{-5}$ or below,
nitrogen-enhanced stars are predicted by our models but there is
currently a dearth of observations of such stars.  From our model, it
is difficult to explain the existence of CEMP stars that have comparable 
[C/Fe] and [N/Fe] with $[\rm{Fe}/\rm{H}]<-3$.  In order to
produce stars with both carbon and nitrogen enriched from our model,
we need hot bottom burning to be less efficient so less carbon is
converted into nitrogen in high-mass AGB stars.  If so, it may be
possible to match those CEMP stars with both carbon and nitrogen
enhanced.

On the other hand, for some stars, there are brief times when both
carbon and nitrogen are simultaneously enhanced, such as the 3$\,\rm
M_\odot$, $Z=10^{-7}$ without overshooting just after its CI episode
but before hot dredge-up.  Similarly, an AGB star can be first
carbon-rich after the onset of third dredge up and then become
nitrogen-rich when hot dredge up kicks in.  If mass is transferred
from the AGB primary to the companion at suitable time, it is possible
for the companion to end up both carbon and nitrogen enhanced.
However, this involves a fine tuning of the period of the system and
is unlikely to make significant amount of carbon and nitrogen enhanced
metal-poor stars.

Finally, we note that the composition of the transferred material may
not necessarily be the same as that of an AGB primary which passes
through its full TP phase. For example, if the star is in a close binary
its life may be truncated if the star fills its Roche lobe or if the
presence of a companion enhances its stellar wind. Therefore the binary
properties of the system and the evolution of the surface abundances
with time are parameters that could affect the observed composition of
the secondary. Moreover, the composition of the transferred material is
also not necessarily the same as what will be observed on the secondary.
Accreted material can become mixed with that of the secondary either
through the deepening of the convective envelope when the star undergoes
first dredge-up, or via thermohaline mixing immediately after accretion.
The consequences of these processes for the surface composition of the
secondary are discussed in detail by \citet{2008MNRAS.389.1828S} and
\citet{Stancliffe09}

\section{The feasibility and problems of deducing the initial mass function}

``What is the initial mass function of metal-poor stars?'' is an
important question that has no conclusive answer at present.  From the
abundances of the observed carbon-enhanced metal-poor stars, one could
theoretically deduce whether they were formed by binary mass transfer
and if so, the initial mass of the primary stars.  As a result, it
ought to be possible to deduce the initial mass function of metal-poor
stars based on the estimated mass of the primary stars.

One important indicator of the primary mass of a particular CEMP star
is the nitrogen abundance, in particular the $[\rm{C}/\rm{N}]$ ratio,
which ranges from $-1$ to around $1.3$ in the observed CEMP stars.  As
mentioned above, a low-mass AGB star that enhances carbon through FDM
or CI has a high $[\rm{C}/\rm{N}]$ ratio unless hot dredge up is
involved.  The $\rm{C}/\rm{N}$ ratio is usually 0 to 2.  On the other
hand, for the high-mass AGB stars, because carbon is converted into
nitrogen, the $\rm{C}/\rm{N}$ ratio is much lower.  The
$[\rm{C}/\rm{N}]$ ratio ranges from around -2 to about 0.
 
Therefore, based on the observed $[\rm{C}/\rm{N}]$ ratio, one might
theoretically deduce whether the primary star is a high-mass or
low-mass AGB stars.  We need more data coming from ongoing surveys such
as SDSS SEGUE, from which seven CEMP stars are recently picked and
discussed in \citet{2008ApJ...678.1351A}.  If enough CEMP stars are
observed, one might deduce the shape of the IMF of low-metallicity
environment by noticing whether low-mass AGB are more numerous or vice
versa.

However we currently cannot match all the observed CEMP stars to
initial primary masses and so a reliable IMF shape cannot be
constructed.  In particular, the high-mass AGB stars cannot fit some
of the CEMP stars with strong enhancement in carbon and nitrogen and
whether we should attribute all those type of stars to high-mass AGB
stars would change the result.  Though we can draw a general
conclusion that a high fraction of CEMP stars with high
$\rm{C}/\rm{N}$ ratio implies more low-mass AGB primary and vice
versa.

On the other hand, our model predicts that both $\rm{C}/\rm{Fe}$ and
the fraction of CEMP stars increase at lower metallicity.  It would be
interesting whether such a trend appears when more metal-poor stars
are observed, particularly CEMP stars with metallicity
$[\rm{Fe}/\rm{H}]<-4$.  If such a trend is not found observationally
or the observed abundances, particularly the C/N ratio, do not fit
well with the surface abundances of AGB models, it could be that there
is a significant shortcoming to current AGB modelling.  Another
possibility is that AGB stars did not form and the initial mass
function peaked at the massive end below a critical metallicity.  If
this is the case, other formation channels are needed to explain those
CEMP stars.

\subsection{The $s$-process elements}

We have not included full nucleosynthesis routines
\citep{2005MNRAS.360..375S}, so abundances of light elements are not
studied.  This nucleosynthesis can also indicate whether
$s$-processing has occurred.  Without it we do not know if stars with
FDM or CI mechanism have any s-process elements.  This could be
important to deduce the mass of the primary because there are two
groups of CEMP stars, one with s-process elements and one without
s-process elements.

\section{Conclusions}

We have investigated the evolutionary properties of extremely-metal
poor AGB stars with $Z=10^{-4}$ to $Z=10^{-8}$.  We show that for
stars with metallicity of $Z=10^{-4}$, the AGB properties are similar
to those of higher metallicity stars, except for the occurrence of hot
third dredge up.  However, as metallicity decreases, flash-driven
mixing or carbon ingestion occurs for low-mass AGB stars.  The highest
metallicity models for which such phenomena occur is $10^{-5}$.  We
propose CI is a mild version of flash driven mixing.

High-mass AGB stars with masses greater than about $4\,\rm M_\odot$
can undergo hot dredge-up.  As a result, the surface $\rm{C}/\rm{N}$
ratio of these stars is lower.  Surface abundances after hot dredge up
depend on mass but not metallicity.  The critical metallicity below
which third dredge up does not occur is $Z=10^{-8}$.  It is possible
for some stars at this metallicity to completely cease thermally
pulsing later in their evolution and end their fate as supernovae type
1.5.  \citep{2008MNRAS.385..301L}.  One model, 5$\,\rm M_\odot$,
$Z=10^{-8}$ without overshooting, is found to have this type of
behaviour.  However it does not occur with overshooting for stars with
$Z=10^{-8}$.  Overshooting also causes the models to behave like a
higher mass models without overshooting.  In general, the mass
boundaries of the different behaviours are shifted by about 1$\,\rm
M_\odot$.

We show that the $\rm{C}/\rm{N}$ ratio of low-mass AGB stars can be as
high as several hundred while the $\rm{C}/\rm{N}$ ratio of high-mass
AGB stars can be less than one.  The variation of carbon and nitrogen
enhancement of CEMP stars could then be due to the different initial
mass of the primary.  However our models here can only explain the
CEMP stars with $\rm{C}/\rm{N}$ much larger unity and those with C/N
around 1/10.  They struggle to explain CEMP stars with similar
enhancements of carbon and nitrogen.  Hot third-dredge up is needed to
convert carbon to nitrogen but it is too efficient in our models,
destroys too much carbon and produces too much nitrogen to fit the
observations.

Theoretically, based on the observed frequency of different
$\rm{C}/\rm{N}$ ratio of CEMP stars at low metallicity, it is possible
to deduce the shape of initial mass function of AGB stars, as stars
with large $\rm{C}/\rm{N}$ ratio are more likely to accrete mass from
low-mass AGB stars.

\section{Acknowledgements}
We thank the referee, Achim Weiss, for his useful comments and suggestions.
HBL thanks KIAA for his fellowship and the IoA for support.  RJS is
funded by the Australian Research Council's Discovery Projects scheme
under grant DP0879472.  CAT thanks Churchill College for his
fellowship.

\bibliography{lau}

\label{lastpage}

\end{document}